\documentclass[sigconf]{acmart}
\renewcommand\footnotetextcopyrightpermission[1]{}

\usepackage{amsmath,amssymb,amsfonts}
\usepackage{algorithmic}
\usepackage{graphicx}
\usepackage{textcomp}
\usepackage{hyperref}       
\hypersetup{hidelinks}
\usepackage{url}            
\usepackage{booktabs}       
\usepackage{amsfonts}       
\usepackage{nicefrac}       
\usepackage{microtype}      
\usepackage{xcolor}         
\usepackage{soul}
\usepackage{caption}
\usepackage{float} 
\usepackage{subfigure}
\usepackage{amsmath}
\usepackage{amsthm}
\usepackage{algorithm}
\usepackage{algorithmic}
\usepackage{bm}
\theoremstyle{definition}
\usepackage{multicol}
\usepackage{multirow}
\usepackage{makecell}
\usepackage{tcolorbox}
\usepackage{lipsum}
\def\BibTeX{{\rm B\kern-.05em{\sc i\kern-.025em b}\kern-.08em
    T\kern-.1667em\lower.7ex\hbox{E}\kern-.125emX}}

\usepackage{xspace}

\theoremstyle{definition}

\AtBeginDocument{%
  \providecommand\BibTeX{{%
    \normalfont B\kern-0.5em{\scshape i\kern-0.25em b}\kern-0.8em\TeX}}}

\setcopyright{acmlicensed}
\copyrightyear{2024}
\acmYear{2024}
\acmDOI{XXXXXXX.XXXXXXX}

\acmConference[ACM SIGKDD]{ACM Knowledge Discovery and Data Mining}{August 25--29}{Barcelona, Spain}
\acmISBN{978-1-4503-XXXX-X/18/06}

\begin{document}

\title{Towards Unifying Feature Interaction Models for Click-Through Rate Prediction}

\begin{abstract}

Modeling feature interactions plays a crucial role in accurately predicting click-through rates (CTR) in advertising systems. 
To capture the intricate patterns of interaction, many existing models employ matrix-factorization techniques to represent features as lower-dimensional embedding vectors, enabling the modeling of interactions as products between these embeddings. 
In this paper, we propose a general framework called \emph{IPA} to systematically unify these models. 
Our framework comprises three key components: the \emph{Interaction Function}, which facilitates feature interaction; the \emph{Layer Pooling}, which constructs higher-level interaction layers; and the \emph{Layer Aggregator}, which combines the outputs of all layers to serve as input for the subsequent classifier.
We demonstrate that most existing models can be categorized within our framework by making specific choices for these three components. 
Through extensive experiments and a dimensional collapse analysis, we evaluate the performance of these choices. 
Furthermore, by leveraging the most powerful components within our framework, we introduce a novel model that achieves competitive results compared to state-of-the-art CTR models.
PFL gets significant GMV lift during online A/B test in Tencent's advertising platform, and has been deployed as the production model in several primary scenarios.
\end{abstract}

\author{Yu Kang}
 \affiliation{%
    \institution{
    Shanghai Jiao Tong University
    \country{}
    }}
 \email{jerryykang@sjtu.edu.cn}

\author{Junwei Pan}
 \affiliation{%
    \institution{Tencent
    \country{}
    }}
 \email{jonaspan@tencent.com}

 \author{Jipeng Jin}
 \affiliation{%
    \institution{
    Shanghai Jiao Tong University
    \country{}
    }}
 \email{jinjipeng@sjtu.edu.cn}

 \author{Shudong Huang}
 \affiliation{%
    \institution{Tencent
    \country{}
    }}
 \email{ericdhuang@tencent.com}

  \author{Xiaofeng Gao}
 \affiliation{%
    \institution{
    Shanghai Jiao Tong University
    \country{}
    }}
 \email{gao-xf@cs.sjtu.edu.cn}

   \author{Lei Xiao}
 \affiliation{%
    \institution{Tencent
    \country{}
    }}
 \email{shawn@tencent.com}



\maketitle

\section{Introduction}

Online advertising has become a billion-dollar business nowadays, with an annual revenue of 225 billion US dollars between 2022 and 2023 (increasing 7.3\% YoY)~\cite{iab-report}.
One of the core problems in this field is to deliver the right ads to the right audiences in a given context. 
Accurately predicting the click-through rate (CTR) is crucial to solving this problem and has attracted significant attention over the past decade~\cite{fm2010,ffm2016, fwfm2018, xdeepfm2018, fibinet2019, dlrm2019, fmfm2021, autoint2019, dcnv22021, finalmlp2023, zhu2023final, guo2023embedding, pan2024ad}.

CTR prediction commonly involves the handling of multi-field categorical data~\cite{multi-field-categorical2016, fwfm2018}, where all features are categorical and sparse. 
The primary challenge lies in effectively capturing interactions between these features, which is particularly difficult due to the extreme sparsity of feature co-occurrence.
To address this challenge, numerous approaches~\cite{fm2010, ffm2016, fwfm2018, fmfm2021, xdeepfm2018, fibinet2019, autoint2019, dcnv22021} have been proposed to explicitly model feature interactions using matrix factorizations or hybrid methods that combine explicit modeling with deep neural networks (DNNs). 

Explicit interaction models, particularly the 2nd-order ones, have well-defined definitions and close-form formulations. This line of models originated from classic Matrix Factorization (MF)~\cite{mf2009, pmf2008}, followed by FM~\cite{fm2010}, FFM~\cite{ffm2016}, FwFM~\cite{fwfm2018}, and FmFM~\cite{fmfm2021}. 
In contrast, models like xDeepFM~\cite{xdeepfm2018} and DCN V2~\cite{dcnv22021} employ a parameter or weight matrix to model higher-order interactions, going beyond the 2nd-order interactions captured by the aforementioned models.
However, it is worth noting that while some attempts have been made to discuss the connections between explicit interaction models~\cite{dcnv22021, ctr-survey-attention2021, aoanet2021}, these efforts have only covered a limited number of models. 
As a result, the differences and relationships between various explicit interaction models remain less well-understood.

Thus, we propose a simple framework, \textit{i.e.}, IPA, to unify existing low- and high-order explicit interaction models for systematic comparisons between them. 
The name of IPA corresponds to its three components:
the \emph{\underbar{I}nteraction Function} which captures the interaction between two terms (or features), 
the \emph{Layer \underbar{P}ooling} which constructs explicit interaction layers based on the prior layers and raw feature embeddings, 
and the \emph{Layer \underbar{A}ggregator} which takes all layers as input, and outputs a representation for the classifier. 
By making specific choices for these three components within the IPA framework, we can represent various CTR models, ranging from simple 2-order interactions models such as FM~\cite{fm2010}, FwFM~\cite{fwfm2018}, FmFM~\cite{fmfm2021} to high-order interaction models such as xDeepFM~\cite{xdeepfm2018} and CIN~\cite{dcnv22021}. 

Furthermore, the IPA framework enables a more granular analysis of structural differences between these models, delving into the comparison of three component choices to identify potential factors influencing model performance, thereby providing a better guidance for model design. To validate the effectiveness of such analysis, we conduct extensive experiments over public datasets and production dataset, drawing comparison between existing CTR models as well as models derived from our framework. These experiments not only identify the key factors behind model performance, enabling our derived models to be as robust as SOTA models, but also elucidate the relationship between component choices and performance from a novel perspective: the phenomenon of Embedding Collapse.
The main contribution of the paper can be summarized as:

\begin{itemize}
    \item We propose a general framework IPA for feature interaction models, which consists of the Interaction Function, the Layer Pooling and the Layer Aggregator. The framework provides more granular way to analyze the structural differences between existing models.
    \item We derive several novel models by new combinations of IPA components. Experiment results demonstrate the effectiveness of these new models on both public datasets and online testing, providing essential findings for model designing.
    \item We present a novel Dimensional Collapse perspective in understanding the evolution of Interaction Functions and deep analysis in the capability of learning data orders.
\end{itemize}

The structure of this paper is as follows: Initially, in section 2, we shall define the three components of our IPA framework, explain the various choices for each of these components as well as real-world applications, and illustrate how the components formulate our framework. Subsequently, section 3 will detail our experiment design, dataset preparation, and our extensive analysis on the proposed research questions. While the following section 4 discusses works related to our field of interest, we shall conclude our research in the final section 5.

\section{The IPA framework}

\begin{figure*}[h!]
    \centering
    \subfigure[The IPA framework.]{\includegraphics[width=0.4\linewidth]{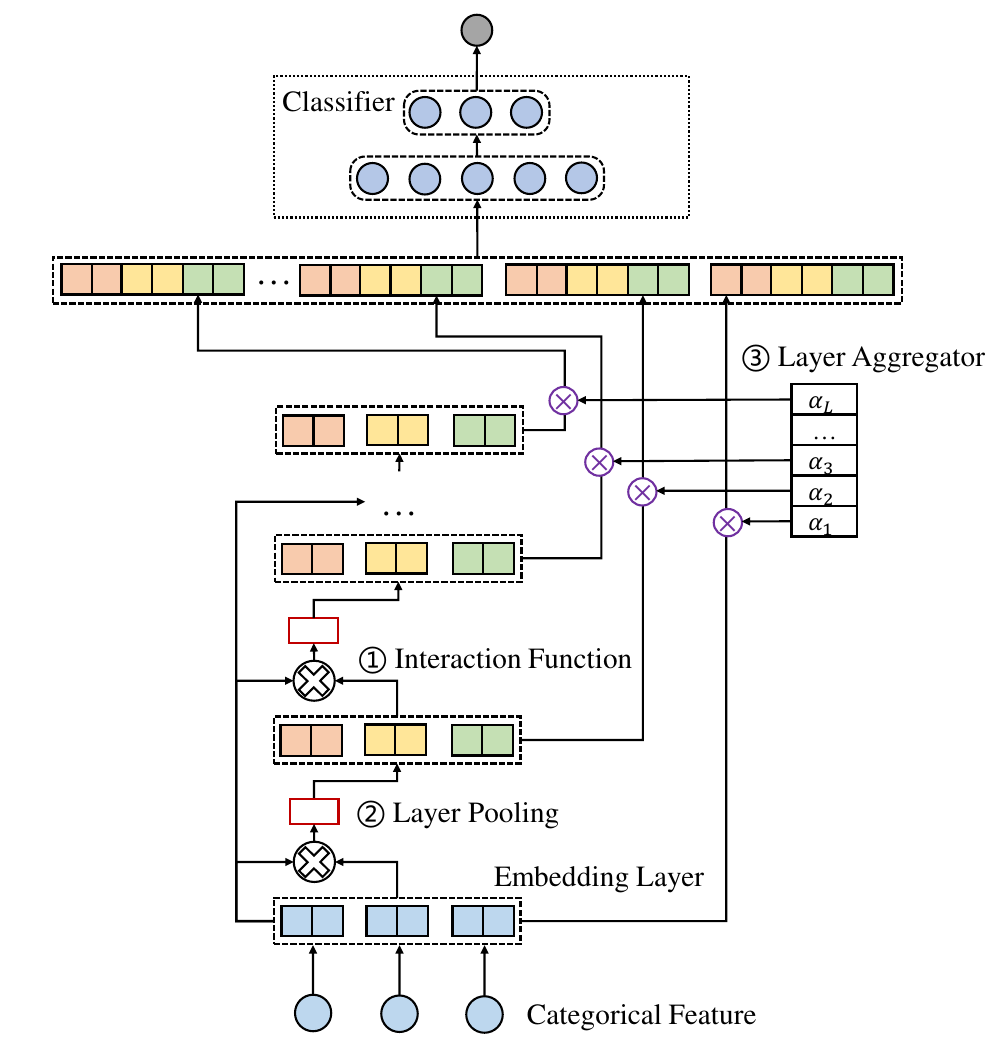}\label{subfig:IPA}}
    \subfigure[Illustration of the components in the IPA framework.]{\includegraphics[width=0.58\linewidth]{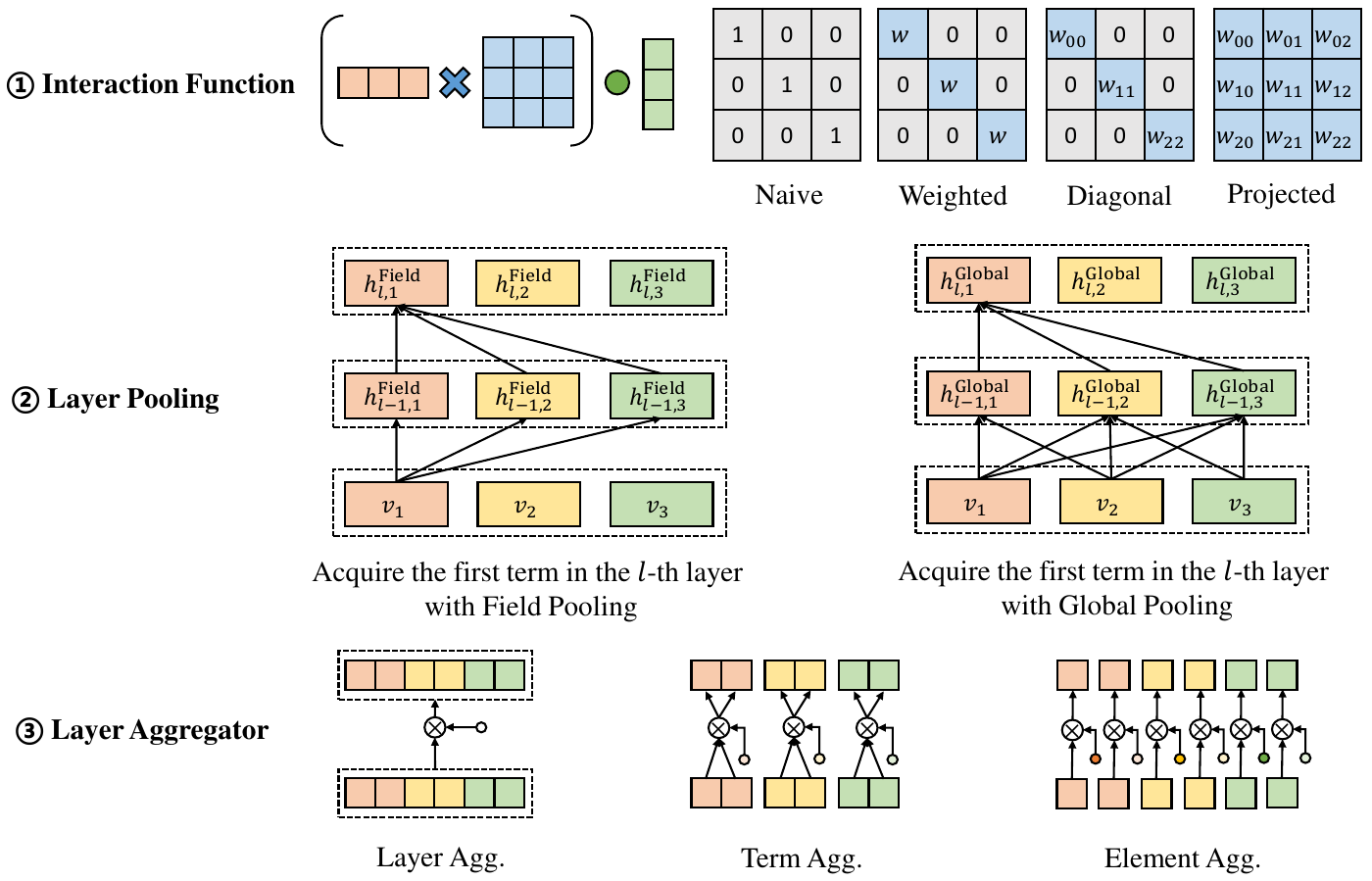}\label{subfig:components}}
    \caption{Illustration of the IPA framework and the common choices of its three components.}
    \label{fig:model}
\end{figure*}

In this section, we present the three modules in our framework: the Interaction Functions, Layer Pooling and Layer Aggregator.

\subsection{Interaction Function}

As the first component of our framework, the Interaction Function is designed to extract information from explicit interaction of the embedding vectors, returning the interaction result in vector form:

\textbf{Definition 1 (Interaction Function)} \textit{Considering two input embeddings $\textbf{t}_i,\textbf{t}_j \in R^K$ , the Interaction Function f is defined as a function mapping $\textbf{t}_i,\textbf{t}_j$ to the interaction term of two these two embeddings:}

\begin{equation}
    \bm{t}_{i,j} = f(\bm{t}_i,\bm{t}_j)  \in R^K
\end{equation}

The input embeddings are not limited to raw embeddings directly obtained from features. When being applied to the embeddings of two raw features $i$ and $j$, the result term represents an explicit second-order interaction of the pair $(i, j)$;
when being used to an interaction embedding, the exact meaning of output vectors becomes complicated, but the overall purpose of capturing high-order interaction patterns never changes.

In our framework, illustrated in the middle part of Figure ~\ref{subfig:IPA}, the Interaction Function serves as the basic unit of feature interaction operations. Utilizing the interaction terms generated by the Interaction Function, the whole CTR model is able to model interaction patterns throughout the training progress.

From our observations, the Interaction Function in many existing models can be formulated by an interaction matrix $\bm{W} \in \mathcal{R}^{K \times K}$:

\begin{equation}
    f(\bm{t}_i, \bm{t}_j, \bm{W}) = (\bm{t}_i^\top \bm{W}) \odot \bm{t}_j^\top
    \label{eq:interaction_function}
\end{equation}

We find most of them equipping one of the following four forms of $\bm{W}_{i,j}$, each corresponding to a form of Interaction Function: 

\begin{table}[h]
    \centering
    \caption{Formulation of various Interaction Functions, where diag() denotes the diagonal matrix.}
    \label{tab:interaction_function}
    \addtolength{\tabcolsep}{-1pt}
    \begin{tabular}{llll}
    \toprule[1.3pt]
    Type & Notation & Matrix & Example \\
    \midrule
    Naive & $\bm{W}^\text{N}$ & $\bm{I}$ & FM~\cite{fm2010}, HOFM~\cite{hofm2016} \\
    Weighted & $\bm{W}^{\text{W}}$ & $\text{diag}(w,\dots,w)$ & FwFM~\cite{fwfm2018}, xDeepFM~\cite{xdeepfm2018}\\
    Diagonal & $\bm{W}^{\text{D}}$ & $\text{diag}(w_1,\dots,w_K)$ & FvFM~\cite{fmfm2021}\\
    Projected & $\bm{W}^{\text{P}}$ & full matrix & FmFM~\cite{fmfm2021}, DCNV2~\cite{dcnv22021}\\
    \bottomrule[1.3pt]
    \end{tabular}
\end{table}

The first part of Fig.~\ref{subfig:components} illustrates the Interaction Function and four forms for the interaction matrix. The yellow elements in the interaction matrix indicate that the parameters are trainable.

\paragraph{Embedding Collapse of Interactions}\label{sec:collapse}
Recent work~\cite{guo2023embedding} reveals that feature interaction on recommendation models leads to the dimensional collapse of embeddings, that is, the embeddings may only be able to span a low-dimensional space.
Inspired by~\cite{jing2022understanding}, which proves that the projection matrix alleviates the dimensional collapse in contrastive learning, we study if the projection matrix in the Interaction Functions can achieve the same goal.
Refer to Sec.~\ref{subsec:exp_collapse} for details.

\subsection{Layer Pooling}

Building layers to capture explicit high-order interactions has been a critical research topic~\cite{dcn2017, xdeepfm2018, autoint2019, dcnv22021}. However, the exponentially increasing number of combinations($\binom{M}{l}$) makes it unwise for models to present all possible interactions. Thus, the interaction terms need to be explored systematically with tractable complexity. 

A widely-employed method of traversal is to generate interaction terms layer by layer corresponding to their order, starting from the first layer which is simply a concatenation of embeddings of all fields. 
The embedding $\bm{t}_n$ of field $n$ is defined as either the embedding of the active feature for a one-hot encoding field, or a pooling over embeddings for all active features for a multi-hot field. 
Formally,

\begin{equation}
    \bm{h}_1 = [\bm{t}_{1}, \dots, \bm{t}_{n}, \dots, \bm{t}_{M}]
    \label{eq:first_layer}
\end{equation}

Now, with the first layer available, the Layer Pooling module is able to construct terms of following layers: 

\textbf{Definition 2 (Layer Pooling)} \textit{Considering the raw embeddings $h_1=[\bm{t_1,..., t_M}]$ and the layer of all $(l-1)$-order terms
 $h_{(l-1)}=[\bm{t_{l-1,1},...,t_{l-1,M}}]$, then the Layer Pooling is defined as the way to generate the interaction terms of order $l$ from all interactions between terms of $h_1$ and $h_{(l-1)}$, that is:}

\begin{equation}
    \begin{aligned}
        \bm{t}_{l,n} = \sum_{m=1}^M \sum_{n=1}^M \alpha_{m,n}f(\bm{t}_n, \bm{t}_{l-1,m})
    \end{aligned}
\end{equation}

$\alpha_{m,n}$ are either 0 or 1 indicating presence of corresponding terms.

In our framework, illustrated in the middle bottom part of Figure ~\ref{subfig:IPA}, the Layer Pooling serves as the overall guidebook of feature interaction operations. When a model needs to capture higher-order interactions, it refers to the Layer Pooling to systematically generate and combine new interaction terms, forming new interaction layers of higher order. Furthermore, to deal with the exponentially increasing number of interactions, two widely employed methods build high-order interaction layers with a fixed number of terms, where each term is a pooling of interactions between terms from prior layers and raw feature embeddings.

\subsubsection{Field-wise Layer Pooling(Field Pooling)}

The field pooling will pool all the interactions that correspond to a specific field.
Specifically, the $l$-th layer consists of $M$ terms, with the $n$-th term $\bm{t}_{l, n}$ defined as a pooling over all interactions between the $n$-th term in the first layer, \textit{i.e.}, the embedding of the $n$-th field ${\bm{t}_n}$, and all terms in the $(l-1)$-th layer, \textit{i.e.}, $\bm{t}_{l-1, m}$. 

\begin{equation}\label{eq:AFT}
\begin{aligned}
    \bm{h}^{\text{Field}}_l &= [\bm{t}_{l,1}^{\text{Field}}, \dots, \bm{t}_{l,n}^{\text{Field}},  \dots, \bm{t}_{l,M}^{\text{Field}}] \\
    \quad\bm{t}_{l,n}^{\text{Field}} &= \sum_{m=1}^M f(\bm{t}_n, \bm{t}_{l-1,m}^{\text{Field}}, \bm{W})
\end{aligned}
\end{equation}

DCN V2 employs Field Pooling and Projected Product to build up layers, \textit{i.e.}, $\bm{h}^{\text{CrossNet}}_{l,n} = \sum_{m=1}^M f( \bm{t}_n, \bm{t}_{l-1,m},\bm{W}^F_{l,n,m})$.

\subsubsection{Global-wise Layer Pooling (Global Pooling)}

The field pooling will globally pool all the interactions, regardless of fields.
Specifically, the $l$-th layer consists of $H_l$ terms, with the $n$-th term defined as a pooling over all interactions between all terms in the first layer, and all terms in the $(l-1)$-th layer. 
\begin{equation}\label{eq:AGT}
\begin{aligned}
    \bm{h}_l^\text{Global} &= [\bm{t}_{l,1}^\text{Global}, \dots, \bm{t}_{l,n}^\text{Global}, \dots, \bm{t}_{l,H_l}^\text{Global}] \\
    \bm{t}_{l,n}^\text{Global} &= \sum_{n=1}^{M}\sum_{m=1}^{H_{l-1}} f(\bm{t}_n, \bm{t}_{l-1,m}^\text{Global},\bm{W})
\end{aligned}
\end{equation}

Global Pooling introduces more redundant interactions than in Field Pooling, since all terms in the same layer are symmetric to each other.
xDeepFM~\cite{xdeepfm2018} employs AGT to construct layers, \textit{i.e.}, $\bm{h}^{\text{CIN}}_{l,n} = \sum_{n=1}^M \sum_{m=1}^M f(\bm{t}_n, \bm{t}_{l-1,m},  \bm{W}^F_{l,n,m})$.
The second part of Fig.~\ref{subfig:components} illustrates the details about how to acquire the first term in the $l$-th layer with both pooling.

\subsection{Layer Aggregator}

While Layer Pooling recursively builds layers of higher order, the complex relationship between input features requires model to learn from interaction of all orders. Thus, after all the interaction layers get constructed, the model should systematically extract information from these sets of interaction terms, combining them into an 1-D vector to feed the following classifiers. Here we define the last component of our framework as the Layer Aggregator:

\textbf{Definition 3 (Layer Aggregator)} \textit{Considering $[h_1,...,h_L]$ as interaction layers of model, then the Layer Aggregator LA is defined as a function aggregating elements of layers into one output vector as input of classifier, that is:}

\begin{equation}
     \bm{r} = LA([\bm{h}_1,...,\bm{h}_L])
\end{equation}

Illustrated in the middle-top part of Figure ~\ref{subfig:IPA}, the Layer Aggregator component takes the interaction layers as input, integrates terms within layers and then aggregates all the layers to form the final output of feature interaction module, serving as the input of following classifier in the model.

Here are some widely employed ways to aggregate layers:

\begin{itemize}
    \item Direct Agg.: Directly link each layer, with No weights at all.
    \item Layer Agg.: Assign a layer-wise weight $\alpha_l$ for each layer.
    \item Term Agg.: Assign a Term-wise weight for each term in each layer, \textit{e.g.}, assign $\alpha_{l, n}$ for the $n$-th term of the $l$-th layer. 
    \item Element Agg.: Assign an Element-wise weight for each element in each layer, \textit{e.g.}, assign $\alpha_{l, n, k}$ for the $k$-th element in the $n$-th term of the $l$-th layer. 
\end{itemize}

The third part of Fig.~\ref{subfig:components} illustrates how to assign weights to a term with Layer, Term and Element Agg., respectively.
Tab.~\ref{tab:connec} summarizes the formulation and number of learnable parameters of different Layer Aggregators.

\begin{table}[htbp]
    \centering
    \caption{Formulation of various aggregators, where $\|\{\cdot\}$ denotes the concatenation function.}
    \label{tab:connec}
    \addtolength{\tabcolsep}{-1pt}
    \begin{tabular}{llll}
    \toprule[1.3pt]
    Abbr. & Weight & \# Param. & Output \\
    \midrule
    Direct & 1 & 0 & $\bm{r} = \sum_{l=1}^L {\|}_{n=1}^M\{\bm{t}_{l,n}\}$ \\
    Layer & $\alpha_{l}$ & $L$ & $\bm{r} = \sum_{l=1}^L \alpha_l \cdot {\|}_{n=1}^M \{\bm{t}_{l,n}\}$\\
    Term & $\alpha_{l, n}$ & $LM$ & $\bm{r} = \sum_{l=1}^L {\|}_{n=1}^M\{\alpha_{l,n} \cdot \bm{t}_{l,n}\}$\\
    Element & $\alpha_{l, n, k}$ & $LMK$ & $\bm{r} = \sum_{l=1}^L {\|}_{n=1}^M\{{\|}_{k=1}^K \{\alpha_{l,n,k} \cdot \bm{t}_{l,n,k}\}\}$\\
    \bottomrule[1.3pt]
    \end{tabular}
\end{table}

\subsection{The Framework}

Now, combining the three components above, we propose our framework \textbf{IPA} as a modularized transformation:

\begin{itemize}
    \item First of all, IPA reads in the embedding vectors as the 0th layer and prepares the Interaction Function.
    \item Then, IPA constructs the interaction layers using its Interaction Function on the built layers guided by its Layer Pooling.
    \item Finally, IPA aggregates all the layers by its Layer Aggregator, creating output vector of desired length for following classifiers.
\end{itemize}

By recognizing the structure of feature interaction module as the combination of three components, the IPA framework can drill into the details of existing CTR models, exploring their similarity and differences on various aspects. To be more specific, we can compare the way models make basic interactions, the way they build up high-order interaction terms, and the way they extract information from their interaction layers. 

Besides, the framework can help us control variables when researching on the effect of single components(like Interaction Function), by simply controlling the other components to be the same.

\subsection{Deriving New Models}

\begin{figure}[h!]
    \centering
    \subfigure[Interaction Functions.]{\includegraphics[width=0.3\linewidth]{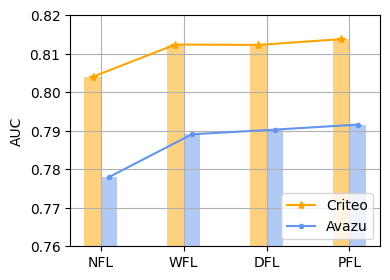}\label{fig:IF_performance}}
    \subfigure[Layer Poolings.]{\includegraphics[width=0.3\linewidth]{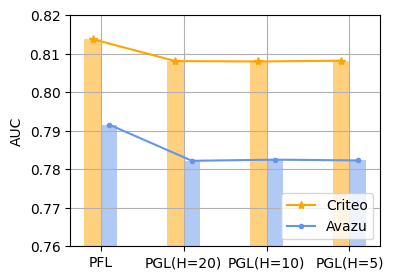}\label{fig:LA_performance}}
    \subfigure[Layer Aggregators.]{\includegraphics[width=0.3\linewidth]{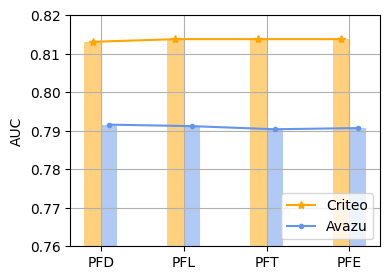}\label{fig:Agg_performance}}
    \caption{Performance of various choices within each component in the IPA framework.}
    \label{fig:components_performance}
\end{figure}

While many existing models can fit into IPA as a combination of three components, there are several new models that can be derived from unexplored combinations.
Specifically, similar to FmFM and DCN V2, we choose Projected Product to capture interactions between terms, 
employ Field Pooling to build up layers and employ Layer Agg. to assign layer-wise weight when aggregating them.
We name it as PFL, which stands for \underbar{P}rojected Product with \underbar{F}ield Pooling and \underbar{L}ayer Agg..
There are two main differences between PFL and DCN V2:
\begin{enumerate}
    \item PFL employs Field Pooling without residual connections, while DCN V2 has residual connections.
    \item PFL employs Layer Agg. to aggregate layers, while DCN V2 uses Direct Agg..
\end{enumerate}

There are many other new models that can be derived in this way, and we present some of the most effective ones in Tab.~\ref{tab:performance}.

\section{Evaluation}
\label{sec:evaluation}

In this section, we aim to answer several research questions related to the performance of components in the Interactor framework.
\begin{itemize}
    \item RQ1: How does each Interaction Function perform, and does a more complicated Interaction Function lead to better performance in the real-world scenario?
    \item RQ2: Regarding the Embedding Collapse, how does Interaction Function influence the extent of collapse?
    \item RQ3: How do the various Layer Poolings perform? What can we infer from them about model design? 
    \item RQ4: How do the various Layer Aggregators perform? What can we learn from choices of this component? 
    \item RQ5: How do existing CTR models evolve so far? How does our framework offer a direction towards a better model?
\end{itemize}

\subsection{Experiment Setup}
\subsubsection{Public Dataset}
We evaluate several state-of-the-art CTR models and some new ones derived from our Interactor framework on two public datasets: Criteo-x1 and Avazu. Both datasets are randomly split into 8:1:1 for training, validation and test.

\begin{itemize}
    \item Criteo-x1~\cite{criteo}. It is the most popular benchmark dataset for CTR prediction, consisting of 45 million click feedback records of display ads. The dataset includes 13 numerical feature fields and 26 categorical feature fields.
    \item Avazu~\cite{avazu}. It includes 10 days of Avazu click-through data, which has 13 feature fields that have disclosed meanings and 9 anonymous categorical fields.
\end{itemize}

\subsubsection{Synthetic Dataset}
We follow~\cite{dcnv22021} to generate data with specified feature interaction orders.
Let $\bm{x} = [x_1, x_2, \dots, x_n]$ be a feature vector of length $n$, and the element $x_i\in \mathcal{R}$ denotes the $i$-th feature of $\bm{x}$. 
$x_i$ is uniformly sampled from interval $[-1, 1]$. 
The monomial $x_1 x_2 \cdots x_O$ is called $O$-order cross-term, representing an $O$-order interaction between features. 
Given $\bm{x}$ of length $n$ and the data order $O$, we generate a label which is the sum of the cross-terms whose orders $\le O$, each with an individual weight.
Specifically, we define $\bm{I}_n = \{i|i\le n, i \in \mathbb{N}^*\}$ as the set of all positive integers less than or equal to $n$, 
and $\Omega_n^O$ as the set of all possible combinations of $O$ distinct integers randomly sampled from $\bm{I}_n$.
The label $y$ is generated by the equation
\begin{equation}
    y = \sum_{i=1}^O\sum_{(\omega_1,  \cdots, \omega_i) \in \Omega_n^i}w_{\omega_1,  \cdots, \omega_i}\prod_{j=1}^i x_{\omega_j} + \epsilon
\end{equation}
where $\omega_i$ is the $i$-th element in the combination, $w_{\omega_1,  \cdots, \omega_i} \sim N(0, 1)$ is the individual weight of a specific cross-term, and $\epsilon \sim N(0, 0.1)$ models the label noise which is ignored in~\cite{dcnv22021}. 

We define the order of a dataset as the maximum order of feature interactions, \textit{i.e.}, $O$. 
We can control the order of a generated synthetic dataset by setting its $O$ manually.

\begin{table*}[h!]
    \centering
    \caption{AUC and Logloss of different models in the IPA framework on Criteo and Avazu Dataset. $O$ stands for orders.}
    \label{tab:performance}
    \addtolength{\tabcolsep}{1pt}
    \begin{tabular}{l|c|c|c|ccc|ccc}
    \toprule[1.3pt]
    \multirow{2}{*}{Model} & \makecell{Interaction \\ Function} & \makecell{Layer \\ Pooling} & \makecell{Layer \\ Aggregator}  & \multicolumn{3}{c}{Criteo} & \multicolumn{3}{|c}{Avazu} \\
    & & &  & $L$ & AUC & Logloss & $O$ & AUC & Logloss \\
    \midrule

    FM & naive & field/global & direct & 2 & 0.8009(2e-4) & 0.4507(4e-4) & 2 & 0.7758(1e-4) & 0.3821(2e-4) \\

    DeepFM & naive & field/global & direct & 2 & 0.8122(1e-4) & 0.4399(2e-4) & 2 & 0.7899(5e-4) & 0.3741(5e-4) \\

    HOFM & naive & field/global & direct & 4 & 0.8040(3e-4) & 0.4479(3e-4) & 5 & 0.7781(8e-4) & 0.3807(6e-4) \\

    FwFM & weighted & field/global & direct & 2 & 0.8095(2e-4) & 0.4423(3e-4) & 2 & 0.7854(4e-4) & 0.3768(4e-4) \\

    xDeepFM & weighted & global & term & 4 & 0.8119(2e-4) & 0.4401(2e-4) & 5 & 0.7897(7e-4) & 0.3741(8e-4) \\

    FvFM & diagonal & field/global & direct & 2 & 0.8103(2e-4) & 0.4415(5e-3) & 2 & 0.7870(3e-4) & 0.3754(4e-4)\\
    
    FmFM & projected & field/global & direct & 2 & 0.8115(3e-4) & 0.4403(3e-4) & 2 & 0.7882(5e-4) & 0.3750(3e-4)\\
    
    FiBiNet & projected & field/global & direct & 2 & 0.8113(2e-4) & 0.4405(2e-4) & 2 & 0.7907(4e-4) & 0.3738(5e-4)\\

    DCN V2 & projected & field' & direct & 4 & \underline{0.8137(3e-4)} & \underline{0.4384(4e-4)} & 5 & \textbf{0.7917(1e-4)} & \textbf{0.3729(4e-4)} \\   

    \midrule
    
    \textcolor{blue}{WFL} & weighted & field & layer & 4 & 0.8124(2e-4) & 0.4394(2e-4) & 5 & 0.7891(3e-4) & 0.3746(5e-4) \\    
    
    \textcolor{blue}{DFL} & diagonal & field & layer & 4 & 0.8123(1e-4) & 0.4395(2e-4)  & 5 & 0.7903(9e-4) & 0.3740(7e-4)  \\

    \textcolor{blue}{PFL} & projected & field & layer & 4 & \textbf{0.8138(3e-4)} & \textbf{0.4381(4e-4)} & 5 & \underline{0.7916(4e-4)} & \underline{0.3731(4e-4)} \\

    \midrule
    
    \textcolor{blue}{PFT} & projected & field & term & 4 & \textbf{0.8138(3e-4)} & \textbf{0.4381(3e-4)} & 5 & 0.7904(4e-4) & 0.3738(6e-4) \\
    
    \textcolor{blue}{PFE} & projected & field & element & 4 & \textbf{0.8138(3e-4)} & \textbf{0.4381(4e-4)} & 5 & 0.7907(2e-4) & 0.3735(3e-4)\\
    
    \textcolor{blue}{PFD} & projected & field & direct & 4 & 0.8131(4e-4) & 0.4388(4e-4) & 5 & 0.7912(5e-4) & \underline{0.3732(4e-4)} \\

    \midrule

    \bottomrule[1.3pt]
    \end{tabular}
\end{table*}

\subsubsection{Baseline Models}

We choose the following CTR models as baselines: FM, FwFM, FvFM, FmFM, DeepFM, FiBiNet, HOFM, xDeepFM and DCN V2. Most of them fit in our IPA framework, and we discuss their components in detail in the first part of Appendix.

\subsubsection{Implementation Details}
Our implementation is based on a public PyTorch library for recommendation models\footnote{https://github.com/rixwew/pytorch-fm}. 
We set the embedding size to 16, the dropout rate to 0.2, and launch early-stopping. 
We use the Adam optimizer with a learning rate of 0.001.
The number of layers $L$ of high-order models are set to 4 and 5 for Criteo-x1 and Avazu, based on the best performance.
Our derived models using PGT have 10 terms in each layer.
All models are trained on a NVIDIA V100 GPU with a batch size of 2048.
We repeat all experiments 3 times and report the average AUC and Logloss performance in Tab.~\ref{tab:performance}. 
The best performance and the comparable performance are denoted in \textbf{bold} and \underline{underlined} fonts, respectively.

\begin{figure*}[htbp]
    \centering
    \subfigure[Information Abundance of fields ordered by cardinality]{\includegraphics[width=0.24\linewidth]{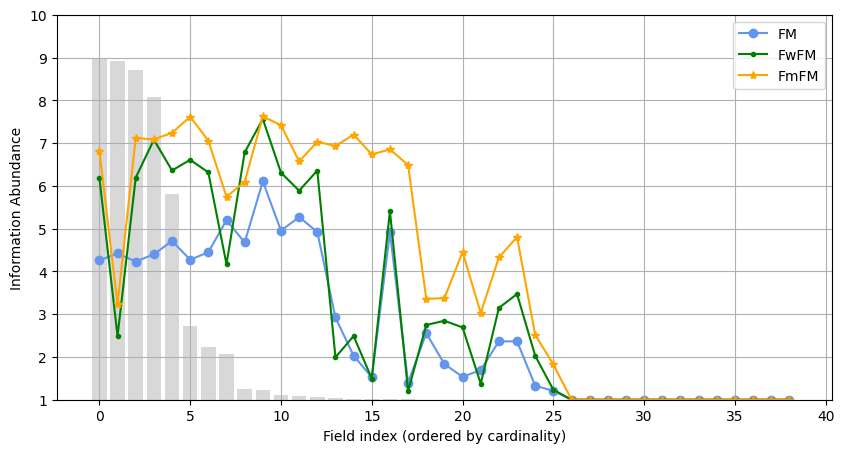}\label{subfig:IA_cardinality}}
    \subfigure[Information Abundance of fields ordered by importance]{\includegraphics[width=0.24\linewidth]{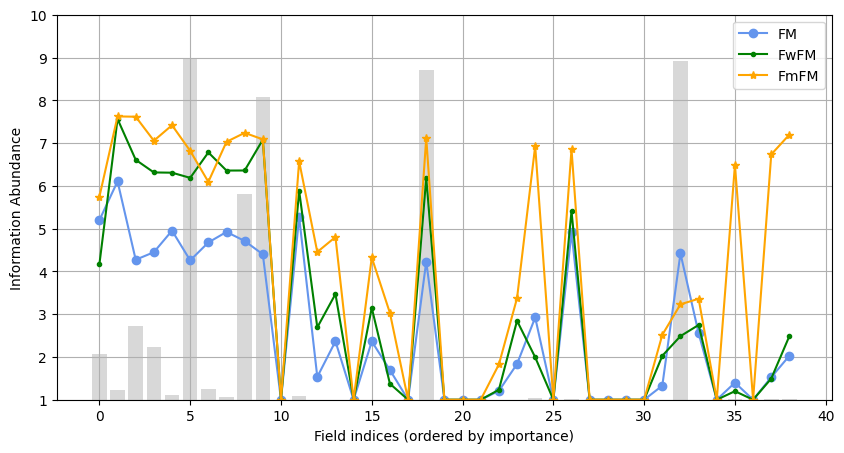}\label{subfig:IA_importance}}
    \subfigure[Singular Sum of fields ordered by cardinality]
    {\includegraphics[width=0.24\linewidth]{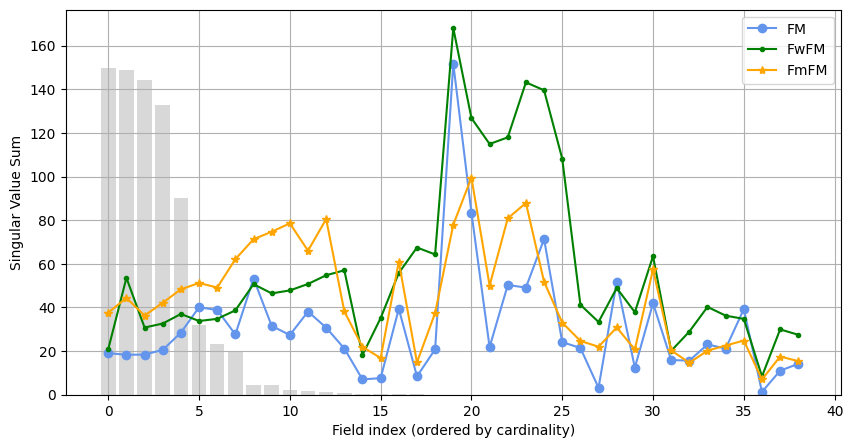}\label{subfig:SV_cardinality}}
    \subfigure[Singular Sum of fields ordered by importance]{\includegraphics[width=0.24\linewidth]{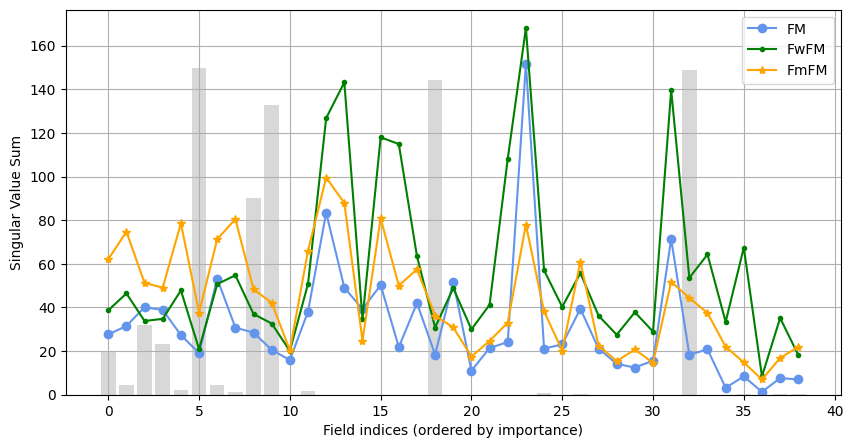}\label{subfig:SV_importance}}
    \caption{Comparison of embedding collapse for 2-order interaction models on the Criteo-x1 dataset. Each of singular value sum and information abundance is plotted for fields aligned in two ways. While fields in (a) and (c) are ordered by field cardinality, those in (b) and (d) are ordered by average feature pair importance in FwFM model. It is clear that FmFM experiences less collapse than FwFM and FM on both high-order fields and important fields, only different in its matrix-projecting Interaction Function.}
    \label{fig:Embedding_Collapse}
\end{figure*}

\subsection{RQ1: Evaluation of Interaction Function}

We compare the models with various Interaction Functions in several settings regarding the Layer Pooling, Layer Aggregator and classifier.
First, among the simplest 2nd order explicit interaction models (Field/Global pooling, direct aggregator, no classifier), \textit{i.e.}, FM, FwFM, FvFM and FmFM, the more complicated the Interaction Function, the better AUC and LogLoss on both datasets.

We compare the models with various Interaction Functions while using the same Layer Pooling (\textit{i.e.}, Field Pooling) and Layer Aggregator (\textit{i.e.}, Layer Agg.). The result is shown in Figure~\ref{fig:IF_performance}.
Across all models (\textit{i.e.}, NFL, WFL, DFL and PFL), PFL achieves the best AUC as well as Logloss. 
Also, comparing the baseline models, the overall AUC and Logloss express a strong trend, as models with more complex Interaction Functions (Projected > Diagonal > Weighted > Naive) generally achieve better AUC and Logloss, indicating the importance of Interaction Function to overall performance.

\begin{tcolorbox}[colback=blue!2!white,leftrule=2.5mm,size=title]
    \emph{%
        Finding 1. Under the same setting of Layer Pooling, aggregation, and classifier, the more complicated the projection matrix (\textit{i.e.}, from identity, scaled identity, diagonal to full matrix) within the feature Interaction Function, the better the results regarding AUC and LogLoss.
    }
\end{tcolorbox}

One possible reason of such trend is that Projected Product learns a powerful matrix projection, $\textit{i.e.}$, $\bm{W}^F$ for each field pair.
Besides, projected product has the greatest number of trainable parameters within the function, further improving its capability to fit on training data, which is likely to connect with model performance.
Additionally, we conduct experiments on the strength of feature interactions for different Interaction Functions(See Appendix). 

\begin{figure}
    \subfigure[Singular Spectrum of high-cardinality fields]{\includegraphics[width=0.9\linewidth]{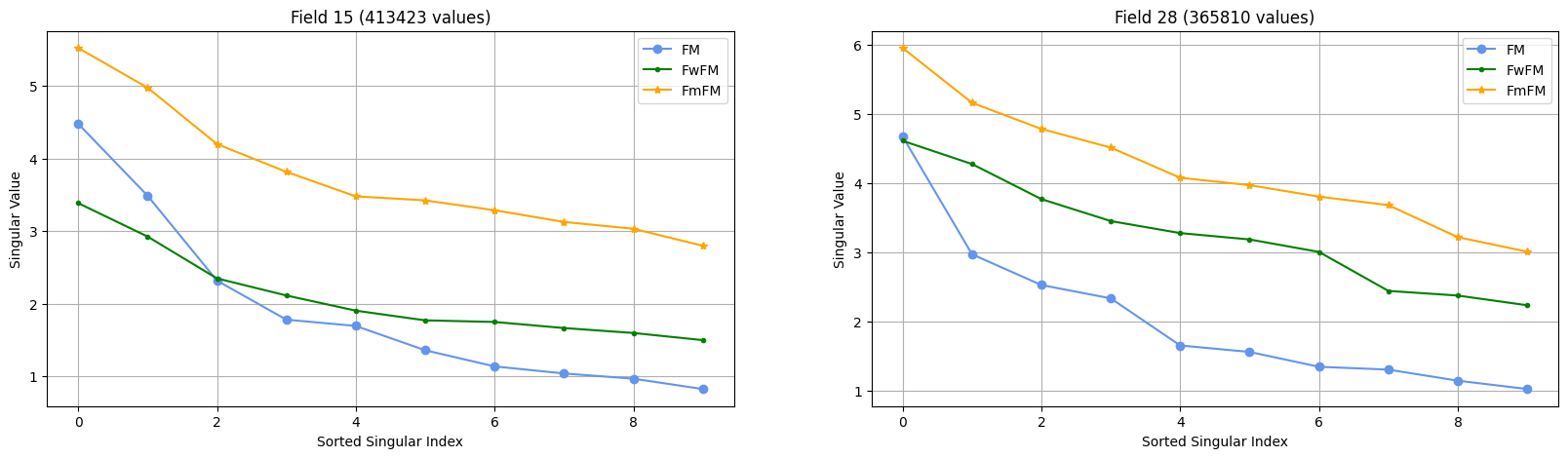}\label{subfig:Fields_cardinality}}
    \subfigure[Singular Spectrum of high-importance fields]{\includegraphics[width=0.9\linewidth]{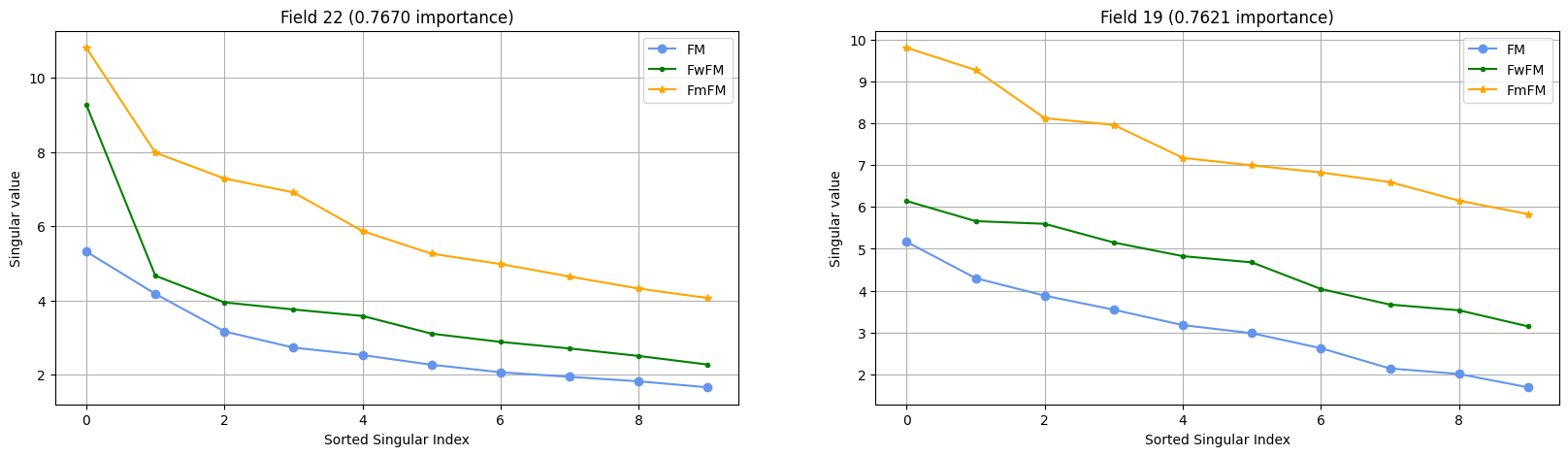}}
    \caption{Field-wise singular value spectrum for 2-order interaction models on the Criteo-x1 dataset. The singular values of representative fields(high-order fields in (a) and high-importance fields in (b)) are ordered and displayed. It is obvious that singular values of FmFM are at the largest level as well as being the most equally distributed among all 2-order models.}
    \label{fig:Singular_Spectrum}
\end{figure}

\subsection{RQ2: Embedding Collapse of Interaction Functions}\label{subsec:exp_collapse}

To explore the extent of Embedding Collapse phenomenon with respect to Interaction Functions, we choose to analyze the singular value sums as well as the information abundance~\cite{guo2023embedding} of feature fields, where the latter one is defined as the singular value sum divided by the largest value.
Different from~\cite{guo2023embedding}, we evaluate these metrics on the embeddings of each feature under the sample distribution rather than the unique feature ID distribution since it reflects the feature frequency.
All the following analysis is based on the Criteo dataset, since its features vary greatly in cardinality to better illustrate the dimensional collapse phenomenon. 

\begin{figure*}[h!]
    \centering
    \subfigure[Trend of layer weights $\alpha_l$]{\includegraphics[width=0.225\linewidth]{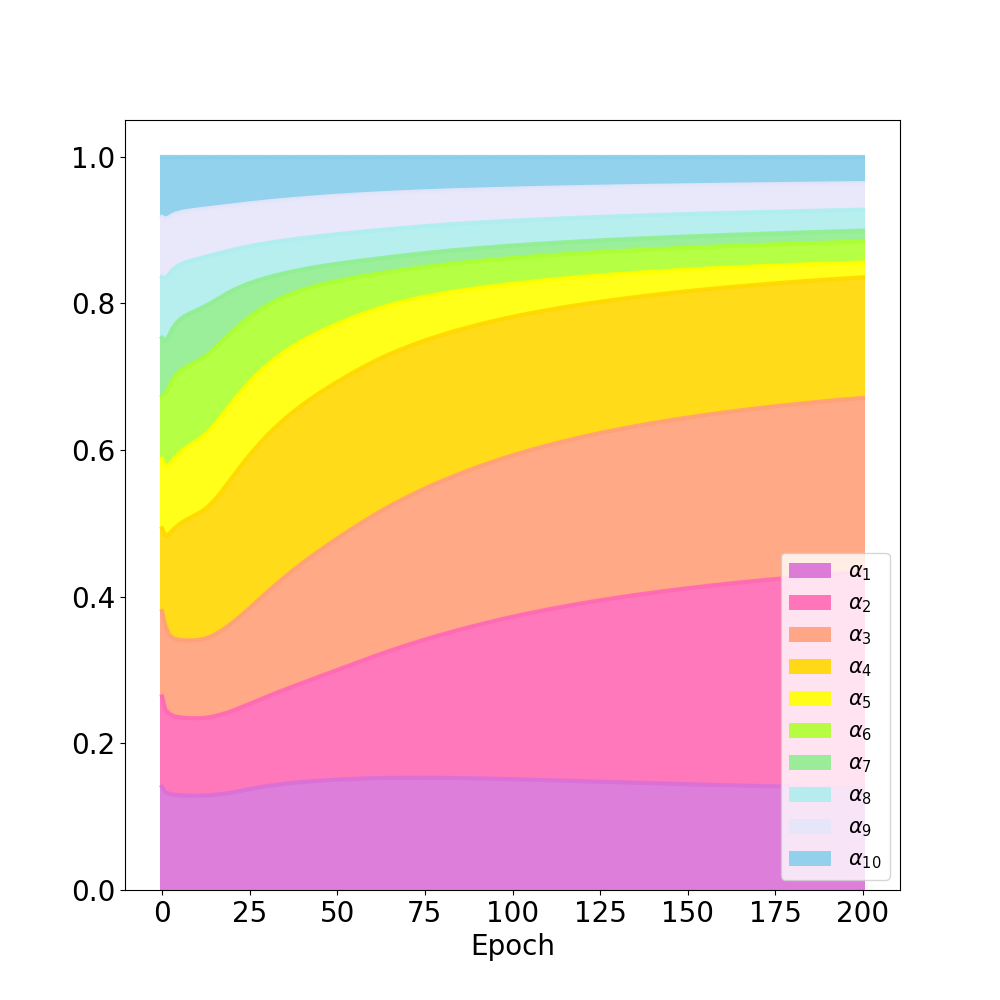}\label{subfig:alpha}}
    \subfigure[Trend of $\alpha_l * \Vert\bm{W}_{l-1}\Vert_\text{F}$]{\includegraphics[width=0.225\linewidth]{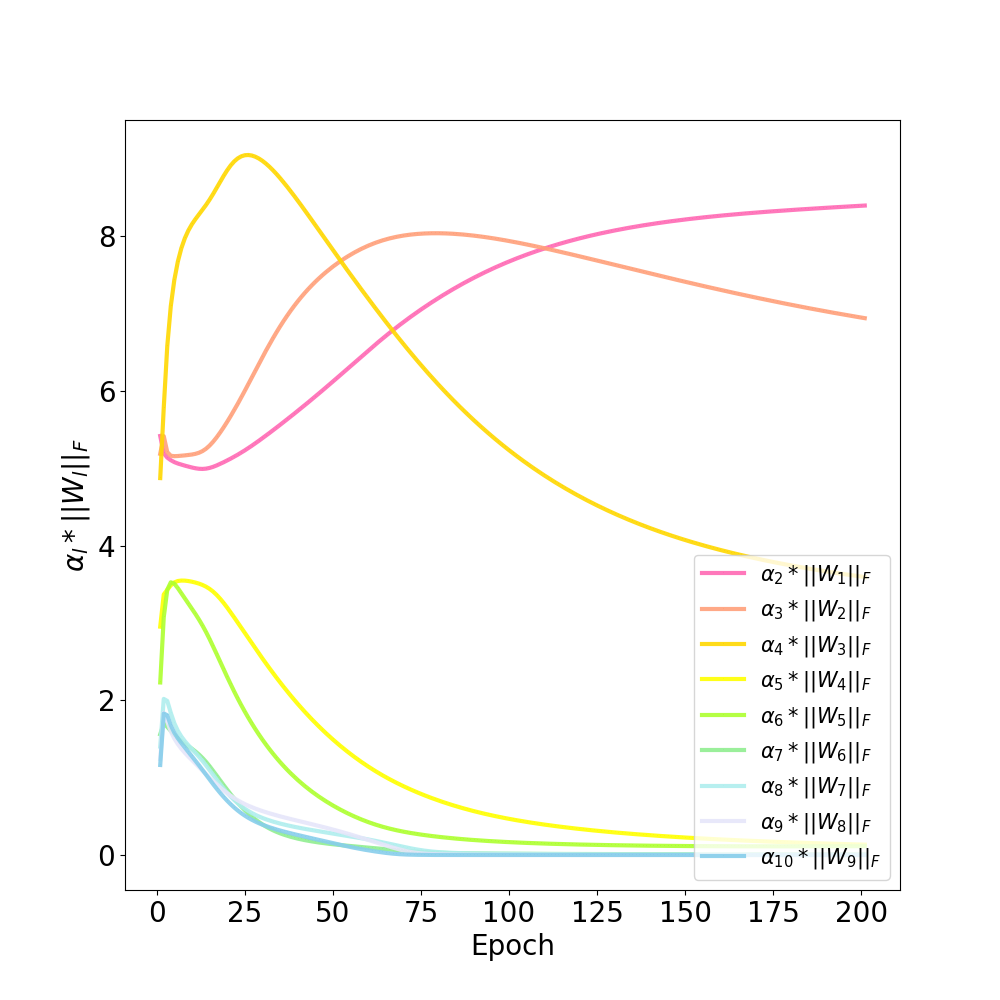}\label{subfig:alpha_W}}
    \subfigure[Performances on data order = 3]{\includegraphics[width=0.225\linewidth]{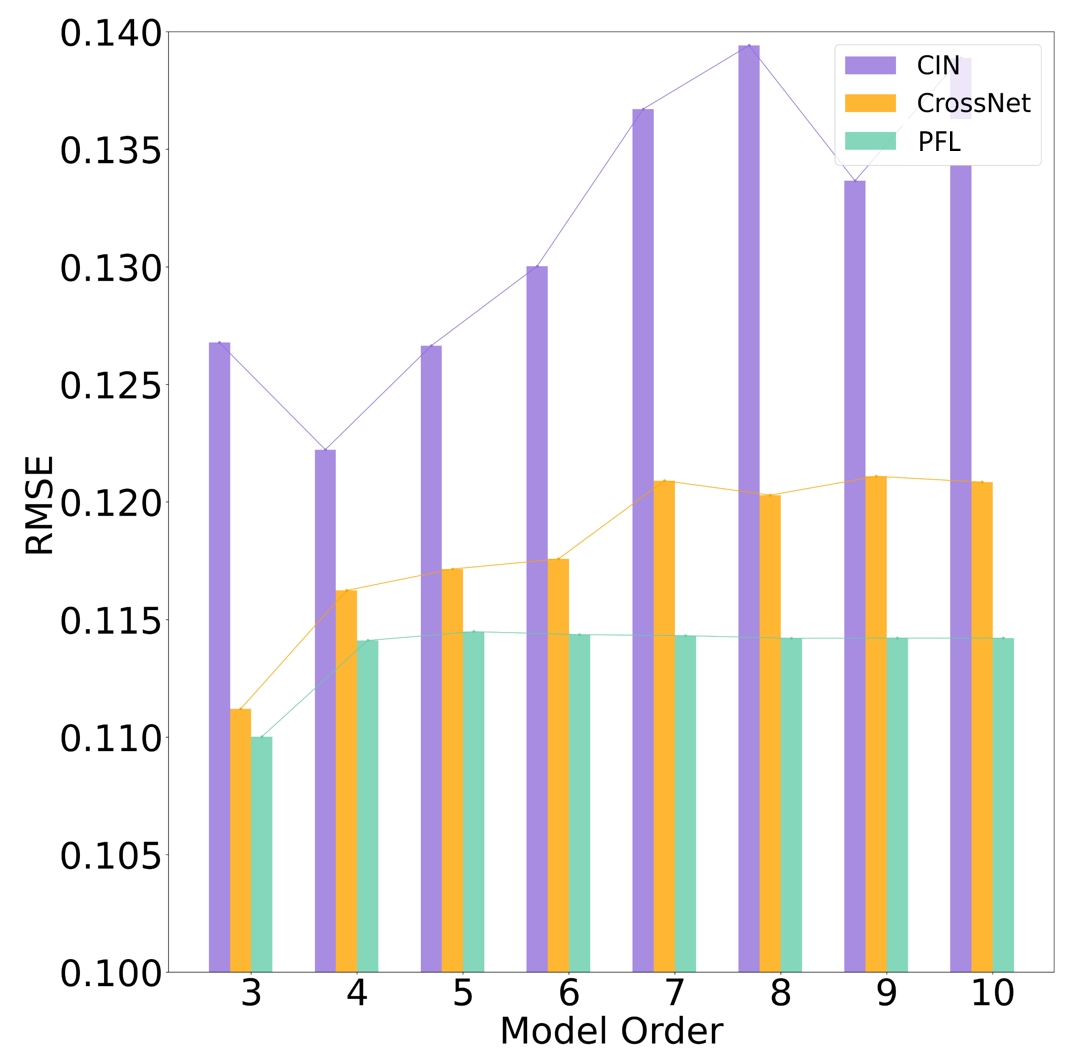}\label{subfig:perf_order_3}}
    \subfigure[Performances on data order = 4]{\includegraphics[width=0.225\linewidth]{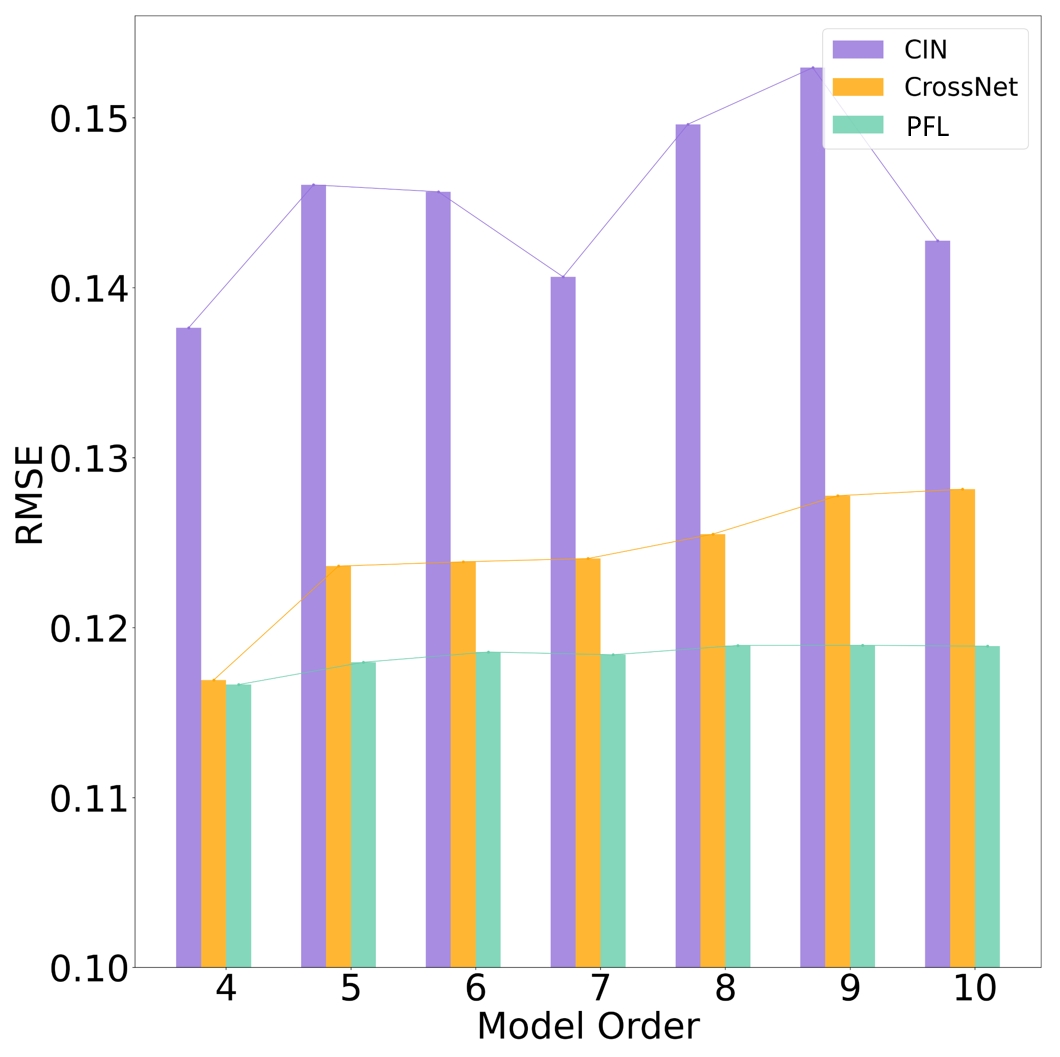}\label{subfig:perf_order_4}}
    \caption{Trends of $\alpha_l$, $\alpha_l * \Vert\bm{W}_{l-1}\Vert_\text{F}$ and model performances in the training process. Our model learns low $\alpha_l$ and $\Vert\bm{W}_{l-1}\Vert_\text{F}$ for extra layers (5-10), obtaining high-level and robust performance even when over-estimating data order.}
    \label{fig:synthetic_experiment}
\end{figure*}

We run experiments on DCNV2 and three representational models: FM deploying Naive Product, FwFM deploying Weighted Product and FmFM deploying Projected Product type, whose only difference lies in the Interaction Functions they take.  
In Figure~\ref{subfig:IA_cardinality} and ~\ref{subfig:SV_cardinality}, we present the singular value sums of sample embeddings in the Criteo-x1 dataset, where feature fields are ordered by their cardinality and average pair importance from the FwFM model.
Comparing the three representational models, we could observe that when the projection matrix of Interaction Function becomes more complex(FM's identity matrix to FwFM's scaled identity matrix to FmFM's full matrix), the singular values of the left-most high-dimensional fields becomes higher and more balanced, obtaining a larger singular sum as well as a larger information abundance; this indicates that FmFM spans the largest space in its high-dimensional field and thus experiencing less amount of Embedding Collapse on this field. Such observation validates that a complex Interaction Function could serve as a buffer for the high-dimensional features, alleviating Embedding Collapse happened during the interaction process with other low-dimensional features. 

Now, as the function complexity connects closely to the extent of Embedding Collapse, we want to make a connection to the model performances mentioned in the previous part. 
In Figure~\ref{subfig:IA_importance} and ~\ref{subfig:SV_importance}, we derive the importance of feature fields from the FwFM model, and present the singular sums as well as information abundance ordered by such importance. 
From the figure, we still observe FmFM to possess the largest singular value sums among the 10 most important feature fields for prediction task, suggesting that models with a complex Interaction Function indeed learns more robust embeddings, which are strong against the Embedding Collapse phenomenon and contribute to a better performance of the model.

Besides, to further discover the pattern of singular values after decomposition, we pick several representative feature fields, either of high cardinality or of high feature importance, and plot the ordered singular values in Figure~\ref{fig:Singular_Spectrum}. From the figure, it is obvious that FmFM possesses the highest and most steady level of singular values among the three 2-order interaction models, not only for high-order fields but also for high-importance fields. This indicates that models with a more complex Interaction Function suffers less from Embedding Collapse and learns a more robust set of embeddings.  

\begin{tcolorbox}[colback=blue!2!white,leftrule=2.5mm,size=title]
    \emph{%
        Finding 2. Among the 2nd-order interaction models, the more complicated the projection matrix (\textit{i.e.}, from identity, scaled identity, diagonal to full matrix) within the feature Interaction Function, the more robust (less collapsed) the learned embeddings regading both information abundance and singular value sum.
    }
\end{tcolorbox}

\subsection{RQ3: Evaluation of Layer Pooling}

To evaluate Layer Pooling, we compare the models with various Layer Poolings, but the same Interaction Function (\textit{i.e.}, Weighted, Diagonal and Projected Product) and the same Layer Aggregator (\textit{i.e.}, Layer Agg.) in Tab.~\ref{tab:performance}.
In addition, we tune the hyper-parameter $H$, \textit{i.e.}, number of terms in each layer in global agg., by training the PGL model with $H=5, 10, 20$, and present the results in Fig.~\ref{fig:LA_performance}. 
Field pooling constantly outperforms Global Pooling in all comparisons, possibly due to that Global Pooling introduces too many redundant interactions, leading to optimization issue caused by co-linearity.

\begin{tcolorbox}[colback=blue!2!white,leftrule=2.5mm,size=title]
    \emph{%
        Finding 3. Under the same setting of feature Interaction Function and Layer Aggregator, models with field pooling constantly outperform those with global pooling.
    }
\end{tcolorbox}

\subsection{RQ4: Evaluation of Layer Aggregator}

Now, in order to analyze the effects of Layer Aggregators,
we equip PFL with four Layer Aggregators defined in the previous section. 
Evaluation in Fig.~\ref{fig:Agg_performance} demonstrates that the aggregators get comparable performance with each other, and PFL itself gets comparable performance with DCN V2 in Tab.~\ref{tab:performance} with a fixed number of layers, \textit{i.e.}, $L=4$ in Criteo-x1 and $L=5$ in Avazu.

However, the potential importance of Layer Aggregators cannot be fully revealed with fixed orders for both models and public datasets. In order to investigate whether the layer weight factor $\alpha_l$ in our Layer Aggregator module can effectively reflect the importance of high-order interactions in a given dataset, the orders of both model and dataset need to be flexible for further analysis.
Thus, we generate a synthetic dataset  consisting of 1 million samples with $O=4$, and then train a derived PFL model with $L=10$. 
Please note that although we use $\alpha_l$ in PFL to explicitly model the importance of layer $l$, the $\bm{W}_l\in \mathcal{R}^{MK\times MK}$ is also order-wise, and $\alpha_l$ may be absorbed by $\bm{W}_l$. 
Therefore, we analyze the effect of $\alpha_l$ as well as its joint influence with $\bm{W}_l$. 
We present their Frobenius norm, i.e., $\Vert \bm{W}_l\Vert_\text{F}$ and $\alpha_l * \Vert\bm{W}_{l-1}\Vert_\text{F}$ in Fig.~\ref{subfig:alpha} and Fig.~\ref{subfig:alpha_W}, respectively.

From these two figures, We observe that as the number of epoch increases, $\alpha_1\sim\alpha_4$ which correspond to the weights of the first four explicit interaction layers, increase significantly, while other order-wise weights, \textit{i.e.}, $\alpha_5\sim\alpha_{10}$, decrease. 
Besides, when considering $\alpha_l * \Vert\bm{W}_{l-1}\Vert_\text{F}$ as a whole, the contribution of $\alpha_l * \Vert\bm{W}_{l-1}\Vert_\text{F}$ is small and can be ignored when $l > 4$, verifying that PFL mainly learns from the most important layers and is capable of learning the order of the data.

Then, to verify whether such learning capability comes from the choice of Layer Aggregator, we generated two synthetic datasets, each with orders $O=3,4$, respectively. 
For each dataset, we train models of CIN (xDeepFM), CrossNet (DCN V2) and PFL with layers in the range of $[O, 10]$. 
In order to make a clear analysis, we remove the MLP classifier from all models and only use the representation for prediction.
The RMSE results are shown in Fig.~\ref{subfig:perf_order_3} and Fig.~\ref{subfig:perf_order_4}.
We do not include models with fewer than $l$ layers, as their performance drops significantly, which is trivial. 

From the performance results, we have some interesting observations regarding the three models. CIN performs badly in all settings: on the one hand, its performance is in general much worse than CrossNet and PFL in all cases with a large margin; on the other hand, it cannot always achieve the best performance when the model order matches the data order. 
For example, when the model order is 3, it achieves the best performance with a model order of 4. Then, compared to CrossNet, our derived model PFL achieves the best performance when the model order matches the data order. 
It manages to make a larger relative lift as more redundant layers are added to the model. 
Besides, no matter how many these redundant layers, the performance of PFL is consistent while the one for CrossNet deteriorates. 
Since PFL differs from DCN V2 in its introduction of $\alpha_l$, the performance gap indicates the success of layer-wise aggregator in filtering redundant layers in PFL, from which we draw our fourth finding.

\begin{tcolorbox}[colback=blue!2!white,leftrule=2.5mm,size=title]
    \emph{%
        Finding 4. Under the same setting of feature Interaction Function and Layer Pooling, models with layer-wise aggregator are capable of learning the order of data, thus outperforming those with direct connections.
    }
\end{tcolorbox}

\subsection{RQ5: Derived models and Evolution of existing models}

Based on the above evaluation, the Projected Product and Field Pooling outperforms other choices while Layer Agg. outperforms other aggregators in the condition of previous two components, we derive a new model by choosing these most powerful choice within each component, and name the new model as PFL, standing for \underbar{P}rojected product with \underbar{F}ield pooling and \underbar{L}ayer agg.
On both datasets, PFL achieves competitive performance with DCN V2.
Besides, many other derived models achieve decent performance. 
For example, WFL achieves competitive performance with xDeepFM, while mainly differing from xDeepFM in the Layer Pooling.

In a nutshell, through all these evaluations, we show that CTR models evolve mainly through: 
1) Employing more powerful Interaction Functions, from Naive Product (\textit{e.g.}, FM, HOFM, DeepFM and NFM), to Weighted Product (\textit{e.g.}, FwFM, PNN and xDeepFM) and Diagonal Product (\textit{e.g.}, FvFM), and finally to Projected Product (\textit{e.g.}, FmFM, FiBiNet, DCN V2 and PFL). 
2) Employing more powerful Layer Poolings, from Global Pooling (\textit{e.g.}, xDeepFM, WGL, DGL, PGL) to Field Pooling (DCN V2 and PFL). 
3) Employing layer-wise Layer Aggregator instead of directly linking everything, in order to adapt to various order of data. 

\subsection{Online A/B Testing}

We developed PFL in one of the world's largest advertising platforms.
The production model employs Heterogeneous Experts with Multi-Embedding architecture~\cite{multi-embedding2023, stem2023, pan2024ad}.
We replace the IPNN expert in the production model with the PFL expert, which models the interactions between more than five hundred user-, ad-, and context-side features.
Multiple embedding tables are learned for all features, each corresponding to one or several experts.

During the two-week 20\% A/B testing, PFL demonstrated promising results, achieving 0.9\%, 3.7\%, 1.2\%, and 2.7\% GMV lift on several vital scenarios, including Moments pCTR, Content and Platform pCTR, and DSP pCTR.
These improvements were statistically significant according to t-tests. 
PFL has been successfully deployed as the production model in the above-mentioned scenarios, leading to a revenue lift by hundreds of millions of dollars per year.

\begin{figure}
    \centering
    \includegraphics[width=0.9\linewidth]{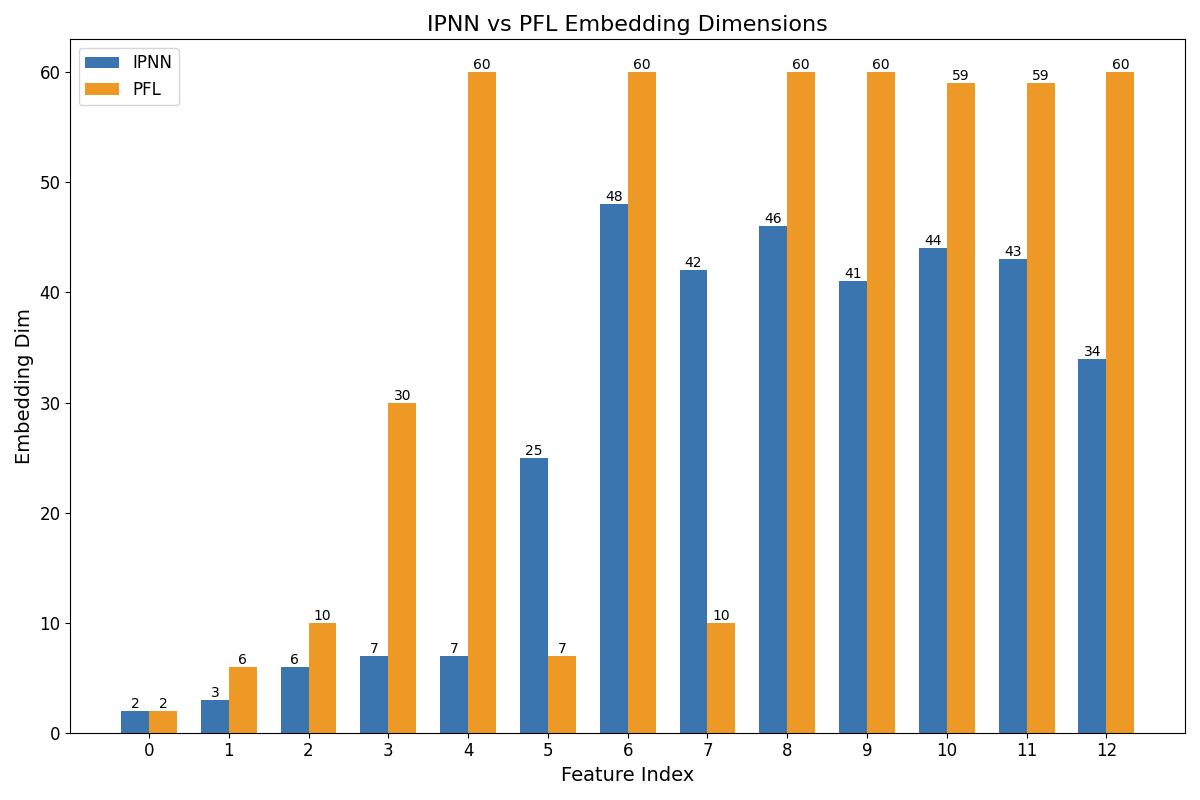}
    \caption{Dimensional Collapse of IPNN v.s. PFL in our system.}
    \label{fig:online_collapse}
\end{figure}

We also study the singular values of the baseline IPNN model and the PFL model.
Specifically, we calculate a 95\%-percentile dimension, that is, how many top singular values can cover 95\% of the total singular values of each feature.
As shown in Fig.~\ref{fig:online_collapse}, our proposed PFL gets a much higher 95\%-percentile dimension on almost all features, especially on those with high cardinalities. 
This validates that PFL can mitigate the dimensional collapse.

\section{Related work}

\label{sec:related}

There has been lots of work in modeling the 2nd-order interactions following factorization machines. 
Early works use Logistic Regression (LR) ~\cite{richardson2007predicting,chapelle2015simple,mcmahan2013ad} or Polynomial-2 (Poly2)~\cite{chang2010training} to learn interactions.
Following matrix factorization~\cite{mf2009, pmf2008}, FM~\cite{fm2010} models interactions as dot product between embeddings. 
FFM~\cite{ffm2016} assigns a field-wise embedding for each feature and only uses the one corresponding the field of the interacted feature in the interaction. 
Based on FM, FwFM~\cite{fwfm2018}, FvFM~\cite{fmfm2021} and FmFM~\cite{fmfm2021} introduce field-pair wise weight, vector and matrix to better model interactions, respectively.
AFM~\cite{afm2017} learns an attentive weight for each feature pair, which takes their embeddings as inputs. 

Many recent work propose to capture explicit high-order ~\cite{fm2010, hofm2016, dcn2017, xdeepfm2018, autoint2019, dcnv22021, li2024dcnv3, eulernet2023, zhu2023final, finalmlp2023}.
For example,~\cite{fm2010} discusses a $d$-way FM to model $d$-order interactions in FM, and HOFM~\cite{hofm2016} presents an efficient algorithm to train it.
Similarly, xDeepFM~\cite{xdeepfm2018}, DCN~\cite{dcn2017}, DCN V2~\cite{dcnv22021} and DCN V3~\cite{li2024dcnv3} also employ variants of traditional matrix factorization techniques to model high-order interactions, while AutoInt~\cite{autoint2019} resorts to a multi-head self-attentive neural network to accomplish this endeavor. 
Additionally, EulerNet~\cite{eulernet2023} learns feature interactions in a complex vector space.
MaskNet~\cite{wang2021MASKNet} introduces multiplicative operations block by block.
FINAL~\cite{zhu2023final} proposes to use multiple FINAL blocks to capture diverse feature interaction patterns.
DCN V3~\cite{li2024dcnv3} uses sub-networks LCN and ECN to capture both low-order and high-order feature interactions

Besides, there is also a lot of work to capture interactions both explicitly via matrix factorization techniques and implicitly via deep neural networks (DNNs).
Wide \& Deep~\cite{wideanddeep2016}, DeepFM~\cite{deepfm2017} and ONN~\cite{onn2020} combine an explicit interaction component (as the wide part) and multiple feed-forward layers (as the deep part) in parallel.
NFM~\cite{nfm2017}, IPNN~\cite{pnn2016}, OPNN~\cite{pnn2016} and FiBiNet~\cite{fibinet2019} employ an explicit interaction component and then add MLPs over it.
FinalMLP~\cite{finalmlp2023} relies on the MLPs and a two-stream structure, achieving amazing performance.
There has been long-standing discussion and debate on the role of MLPs in recommendations~\cite{latentcross2018, dcnv22021, ncfrevisited2020, reenvision2021, tin2024, finalmlp2023, feng2024long}.

There are several works to benchmark~\cite{fuxictr2020, bars2022, recbole[1.0], recbole[2.0], recbole[1.2.0]} and summarize existing CTR models.
For example, AOANet~\cite{aoanet2021} decomposes models into three phases, namely Projection, Interaction, and Fusion.
DCN V2~\cite{dcnv22021} compares the explicit Interaction Functions.

\section{Conclusion}

In this paper, we present a unified framework, \textit{i.e.}, IPA, for explicit interaction click-through rate models to systematically analyze and compare the existing CTR models in Recommender Systems.
With the assistance of IPA framework, we manage to conduct more granular research on the performance of these models, analyzing the effect of three components in extensive experiments and making several interesting discoveries about model design.
Inspired by the framework analysis, we successfully derived a new model that achieves competitive performance with SOTA models and was successfully deployed to Tencent's advertising platform.

\clearpage

\bibliographystyle{ACM-Reference-Format}
\bibliography{6.references}

\clearpage
\appendix
\section{Appendix}

\subsection{Complexity Analysis}
We summarize the space and time complexity of different Layer Poolings and Interaction Functions in Tab.~\ref{tab:complexity_table}.
We ignore the Layer Aggregator since its complexity is negligible.
Please note that for space complexity, we only analyze the additional parameters introduced by modeling feature interactions, neglecting the parameters of feature 
Please refer to~\cite{fm2010} for the proof of the time complexity of the Naive Product.

\begin{table}
    \centering
    \caption{Summary of space and time complexity analysis of all possible architectures under the IPA framework.}
    \label{tab:complexity_table}
    \addtolength{\tabcolsep}{0pt}
    \begin{tabular}{llll}
    \toprule[1.3pt]
    \makecell{Layer\\ Pooling} & \makecell{Interaction\\Function} & Space complexity & Time complexity\\
    \midrule
    Field  &Naive & $O(1)$ & $O(LMK)$ \\ 
    Field & Weighted & $O(LM(M-1))$ & $O(LM^2K^2)$ \\ 
    Field & Diagonal & $O(LMK(M-1))$ & $O(LM^2K^2)$ \\ 
    Field & Projected & $O(LMK^2(M-1))$ & $O(LM^2K^2)$ \\
    \midrule
    Global &Naive & $O(1)$ & $O(LHMK)$ \\ 
    Global & Weighted & $O(LHM(M-1))$ & $O(LHM^2K^2)$ \\ 
    Global & Diagonal & $O(LHMK(M-1))$ & $O(LHM^2K^2)$ \\ 
    Global & Projected & $O(LHMK^2(M-1))$ & $O(LHM^2K^2)$\\
    \bottomrule[1.3pt]
    \end{tabular}
\end{table}

\subsection{CTR models within IPA framework}

In this part, we fit several existing CTR models in the IPA framework, analyzing their interaction process and present them in the style of component choice triplets.

\subsubsection{FM, FwFM, FvFM and FmFM}

Each of these models employs one of the four Interaction Function mentioned above, \textit{i.e.}, FM employs Naive Product, whilst FwFM, FvFM and FmFM employ Weighted, Diagonal and Projected Product, respectively.
Regarding Layer Pooling, each of these models can be regarded as using either Field or Global Pooling to build up its only explicit interaction layer, \textit{i.e.}, the 2nd layer.

From an Global Pooling perspective, each of them has only one term in the 2nd layer, which pools all 2nd-order interactions between all fields. 
Formally,
\begin{equation}
    \begin{aligned}
        \bm{h}_{2} &= [\bm{t}_2^\text{Global}] \\
        \bm{t}_2^\text{Global} &= \sum_{i=1}^M \sum_{j=1}^M f(\bm{t}_i, \bm{t}_j, \bm{W}_{i,j})    \end{aligned}
\end{equation}

From an Field Pooling perspective, the 2nd layer is formulated as:
\begin{equation}
    \begin{aligned}
        \bm{h}_{2} &= [\bm{t}_{2,1}^\text{Field}, \cdots, \bm{t}_{2,i}^\text{Field Pooling}, \dots, \bm{t}_{2,M}^\text{Field}] \\
        \bm{t}_{2,i}^\text{Field} &= \sum_{j=1}^M f(\bm{t}_j, \bm{t}_i, \bm{W}_{i,j})
    \end{aligned}
\end{equation}

All of the models use Direct Agg. as the Layer Aggregator since they do not have any layer-specific weights.
Finally, all these methods choose sum pooling as the classifier.
In summary, FM corresponds to a code of NFD or NGD, standing for \underbar{N}aive Product with \underbar{F}ield Pooling (or \underbar{G}lobal Pooling)  and \underbar{D}irect Agg.,
FwFM corresponds to WFD or WGD, FvFM corresponds to DFD or DGD and FmFM corresponds to PFD or PGD.

\subsubsection{FiBiNet}

There are three feature interaction types in FiBiNet~\cite{fibinet2019}, \textit{i.e.}, Field-All Type, Field-Each Type and Field-Interaction Type. 
All of these types belong to Projected Product, with their definitions slightly different from that of $\bm{W}^P$.
FiBiNet also employs either Global Pooling or Field Pooling to build up layers as it consists of only 2nd-order interactions, and uses Direct Agg. to aggregate layers.
FiBiNet is represented by a code of PFD or PGD, standing for \underbar{P}rojected Product with \underbar{F}ield(or \underbar{G}lobal) Pooling and \underbar{D}irect Agg..

\subsubsection{xDeepFM}

The compressed interaction network (CIN) in xDeepFM~\cite{xdeepfm2018} has the following formula:
\begin{equation}
    \bm{t}_{l,n} = \sum_{n'=1}^{h_{H_l}}\sum_{m=1}^M \bm{W}_{l,n,n',m} (\bm{t}_{l-1,n'} \odot \bm{t}_{m})
    \label{eq:xdeepfm}
\end{equation}
where Eq.~\ref{eq:xdeepfm} itself corresponds to Global Pooling. 
$\bm{W}_{l,n,n',m}$ denotes the pair-wise weight between the $n'$-th term in the prior $l-1$-th layer and the $m$-th field embedding for the $n$-th global-wise term 
This makes $\bm{W}_{l,n,n',m} (\bm{t}_{l-1,n} \odot \bm{t}_{m})$ correspond to Weighted Product.
The $\bm{w}^o$ in CIN for binary classification (as shown in Eq.(8) of paper~\cite{xdeepfm2018}) corresponds to the Term-wise weights of Agg-T.
CIN is represented by a code of WGT, standing for \underbar{W}eighted with \underbar{G}lobal and \underbar{T}erm.

\subsubsection{DCN V2}

The cross network (CrossNet) in DCN V2~\cite{dcnv22021} has the following formula:
\begin{equation}
    \bm{t}_{l,n} = \sum_{m=1}^M  ((\bm{t}_{l-1,n}^\top  \bm{W}_{l,n,m} ) \odot \bm{t}_{m}) + \bm{t}_{l-1,n}
    \label{eq:crossnet}
\end{equation}
where Eq.~\ref{eq:crossnet} corresponds to Field Pooling with residual connections, denoted as Field Pooling'. 
$\bm{W}_{l,n} \in \mathcal{R}^{K \times K}$ denotes the pair-wise matrix between the $n$-th field-wise term in the $l-1$-th layer and the $m$-th field, making $(\bm{t}_{l-1,n}  \bm{W}_{l,n, m} ) \odot \bm{t}_{m})$ correspond to Projected Product.
DCN V2 simply adds MLPs upon the last layer of CrossNet without explicit layer-wise weights, which is represented by a code of PFD, standing for \underbar{P}rojected' with \underbar{F}ield and \underbar{D}irect.

The Interaction Function of CrossNet (DCN) and CrossNet-V2 (DCN-V2) are in the form of $\bm{W}^{\text{D}}$ and $\bm{W}^{\text{P}}$, respectively. 
But their layer definition is different from Field Pooling in that the $l$-th layer of two models contains all interactions from the first order to the $l$-th order, so that they only employ the last layer's output in classification. 
CrossNet chooses Field Pooling to construct layers in the sense that the $i$-th term in the $l$-th layer contains all $l$-order interactions regarding field $i$.
Additionaly, CrossNet incorporates a residual connection linking the $i$-th term in $l$-th with that of the prior $l-1$-th layer, involving all lower-order interactions into the $i$-th term.
On the contrary, the $l$-th layer in Field Pooling consists of all $l$-th order interactions. Therefore, IPA employ a Layer Aggregator to combine all layer outputs before prediction.

\end{document}